\documentclass[twocolumn,11pt]{article}
\begin{document}

\title{\bf Tunable quantum interference:\\
How to make morning, noon, and afternoon states
}

\author{Holger F. Hofmann and Takafumi Ono\\
Graduate School of Advanced Sciences of Matter, Hiroshima University\\ Kagamiyama 1-3-1, Higashi Hiroshima 739-8530, Japan
}

\date{}

\maketitle

\abstract{We show that the $N$-photon states produced by
interference between laser light and downconverted
light at the input of a two path interferometer
can be characterized by a single tuning parameter
that describes a transition from phase squeezing to
nearly maximal path entanglement and back.
The quantum states are visualized on a sphere
using the analogy between $N$-photon interference
and the spin-$N/2$ algebra.
}

\section{Introduction}
The sensitivity of quantum phase measurements is limited by the
quantum fluctuations in the two mode $N$-photon statistics of the
light field states used to probe the phase shift \cite{Gio04}.
If coherent input light is used, the sensitivity is limited by the
shot noise in the seemingly random photon detection events, resulting
in the standard quantum limit of $\delta \phi^2=1/N$. However,
quantum coherence can decrease this phase estimation error up to
the fundamental Heisenberg limit of $\delta \phi^2=1/N^2$.
The $N$-photon state that achieves this maximal phase sensitivity
is the superposition state of the state where all photons are in
one arm of the interferometer and the state where all photons are
in the other arm,
\begin{equation}
\mid \mbox{NOON} \rangle = \frac{1}{\sqrt{2}}\left(
\mid N;0 \rangle + \mid 0;N \rangle
\right).
\end{equation}
In recognition of the technological challenges involved in trying
to realize such states for high photon numbers, Jonathan Dowling
has dubbed these states `high NOON' state, and by now, these states
are commonly referred to as noon states in the literature
\cite{Bow04}.

There have been numerous proposals for the generation of noon states
\cite{noonprop}, leading up to the experimental generation of three
and four photon states showing the expected $N$-photon coherence
\cite{noonexp}. Unfortunately, the rather inefficient post-selection
methods used in these experiments result in low visibilities and
make it difficult to achieve even higher photon numbers. However,
help may be on the way. As we recently discovered, high fidelity
noon states can be generated by simply interfering downconverted
photon pairs and laser light \cite{Hof07}. Experimentally, this
technique has been pioneered by Lu and Ou, and its application to
three photon noon state generation was proposed by Shafiei and
coworkers from the same group. However, it was thought that the
generation of higher photon number noon states is not possible with
this method, because a fidelity of 100 \% can only be
obtained if additional non-linear elements are used \cite{QI}.
In fact, the quantum interference between laser light and
downconverted light generates a slightly squeezed noon state,
indicating that the non-classical states generated by this method
combine squeezing effects with multi-photon coherences.

In our recent research, we have studied the continuous transition
from gradual phase squeezing when most photons originate from the
laser \cite{Ono07} to multi-photon quantum coherences resulting in
a maximal noon state fidelity of about 94\% when the average number
of photons from the laser and from the down-conversion is $N/2$ each
\cite{Hof07}.
In the following, we illustrate the complete transition from
squeezing to noon state and back on the sphere defined by the
spin-$N/2$ algebra of $N$-photon two mode states. It turns out that
the resulting images suggest a new motivation for the noon state
terminology, since the noon state is illustrated by a superposition
of a state at the zenith of the upper half of the sphere and its
mirror image in the lower half. Consequently, we can extend the
terminology to include dawn states (where non-classicality begins
as a slight elongation of a state at the `eastern' horizon),
morning states (where the squeezed state has separated from the
horizon into a superposition within the `eastern' half of the
sphere), afternoon states (where the states have passed the zenith
and moved to the `western' half of the sphere), and evening states
(where the states meet again at the `western' horizon and
non-classicality recedes).

\section{Analogy between $N$-photon interference and spin-$N/2$ rotations}

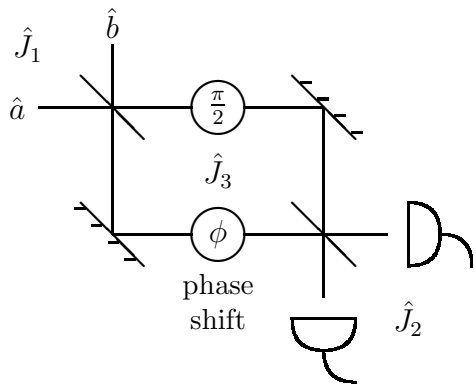
\begin{figure}[th]
\setlength{\unitlength}{0.8pt}
\begin{picture}(300,210)
\thicklines
\put(50,160){\makebox(20,20){\large $\hat{J}_1$}}
\put(90,170){\makebox(20,20){\large $\hat{b}$}}
\put(100,170){\line(0,-1){90}}
\put(45,130){\makebox(20,20){\large $\hat{a}$}}
\put(65,140){\line(1,0){73}}
\put(150,140){\circle{24}}
\put(142,132){\makebox(16,16){\large $\frac{\pi}{2}$}}
\put(162,140){\line(1,0){38}}
\put(85,155){\line(1,-1){30}}
\put(185,95){\line(1,-1){30}}
\put(185,155){\line(1,-1){30}}
\put(189,152){\line(1,0){5}}
\put(197,144){\line(1,0){5}}
\put(205,136){\line(1,0){5}}
\put(213,128){\line(1,0){5}}
\put(85,95){\line(1,-1){30}}
\put(87,92){\line(-1,0){5}}
\put(95,84){\line(-1,0){5}}
\put(103,76){\line(-1,0){5}}
\put(111,68){\line(-1,0){5}}
\put(100,80){\line(1,0){38}}
\put(150,80){\circle{24}}
\put(142,72){\makebox(16,16){\large  $\phi$}}
\put(135,50){\makebox(30,10){phase}}
\put(135,35){\makebox(30,10){shift}}
\put(162,80){\line(1,0){68}}
\put(230,80){\line(-2,1){10}}
\put(230,80){\line(-2,-1){10}}
\put(200,140){\line(0,-1){90}}
\put(200,50){\line(1,2){5}}
\put(200,50){\line(-1,2){5}}

\put(140,100){\makebox(20,20){\large $\hat{J}_3$}}

\put(185,40){\line(1,0){30}}
\bezier{100}(185,40)(185,25)(200,25)
\bezier{100}(200,25)(215,25)(215,40)
\bezier{100}(200,25)(200,10)(215,10)
\put(240,65){\line(0,1){30}}
\bezier{100}(240,65)(255,65)(255,80)
\bezier{100}(255,80)(255,95)(240,95)
\bezier{100}(255,80)(270,80)(270,65)

\put(230,30){\makebox(20,20){\large $\hat{J}_2$}}

\end{picture}
\caption{\label{MZ}
Illustration of two path interferometry using a Mach-Zehnder
interferometer. The components of the Schwinger representation
correspond to half the photon number differences at the input
($\hat{J}_1$), at the output for $\phi=0$ ($\hat{J}_2$), and
inside the interferometer ($\hat{J}_3$).}
\end{figure}

Optical quantum phase measurements can be realized using a two
path interferometer such as the Mach-Zehnder interferometer
shown in fig. \ref{MZ}. An $N$-photon state (or the $N$-photon
component of an arbitrary field state) can then be expressed in
terms of the $N+1$ level system defined by the two mode Fock
states $\mid N-n; n \rangle$. This Hilbert space is
equivalent to that of a spin-$N/2$ system, and the corresponding
spin components can be expressed in terms of the input modes
$\hat{a}$ and $\hat{b}$ using the Schwinger representation,
\begin{eqnarray}
\hat{J}_1 &=& \hspace{0.3cm}
\frac{1}{2}(\hat{a}^{\dagger}\hat{a}-\hat{b}^{\dagger}\hat{b})
\nonumber \\
\hat{J}_2 &=& \hspace{0.3cm}
\frac{1}{2}(\hat{a}^{\dagger}\hat{b}+\hat{a}\hat{b}^{\dagger})
\nonumber \\
\hat{J}_3 &=&
-\frac{i}{2}(\hat{a}^{\dagger}\hat{b}-\hat{a}\hat{b}^{\dagger}).
\end{eqnarray}
As indicated in fig. \ref{MZ}, the three orthogonal components
of this vector can be interpreted as half of the photon number
differences between the input modes ($\hat{J}_1$), between the
output modes at $\phi=0$ ($\hat{J}_2$), and between the paths of
the interferometer ($\hat{J}_3$). A phase shift is then equal to
a rotation around the $J_3$-axis, given by the unitary transformation
$\hat{U}(\phi)=\exp(-i \phi \hat{J}_3)$.

\section{Interference between coherent laser light and
down-converted light}

As Ou and coworkers have pointed out \cite{QI}, it is possible to
generate $N$-photon quantum interferences by mixing coherent laser
light and down-converted light at a beam splitter because the
output measurement does not distinguish between photons from the
laser and photons from the down-conversion. In their theoretical
treatment, Ou and coworkers used the Fock state representations to
derive the relation between the input and the output coherences.
However, we found that a more compact representation of the
coherences can be obtained by using the operator relations that
characterize the coherent state $\mid \alpha \rangle$ in mode
$\hat{a}$ and the single mode down-converted state $\mid \gamma \rangle$ in mode $\hat{b}$,
\begin{eqnarray}
\hat{a} \mid \alpha \rangle &=& \alpha \mid \alpha \rangle
\nonumber \\
\hat{b} \mid \gamma \rangle &=& - \gamma \hat{b}^\dagger
\mid \gamma \rangle.
\end{eqnarray}
The quantum coherence of the $N$-photon component
$\mid \eta \rangle$ of the two mode state can then be characterized
by photon number preserving combinations of creation and annihilation
operators for the two modes \cite{Hof07}.
The most compact relation reads
\begin{eqnarray}
\hat{a}^\dagger \hat{b} \mid \eta \rangle &=&
\eta \frac{\hat{a}^\dagger \hat{a}}{N} \hat{b}^\dagger \hat{a}
\mid \eta \rangle
\nonumber \\ \mbox{where} &&
\eta=\frac{N \gamma}{\alpha^2}.
\label{eq:op}
\end{eqnarray}
Thus the two mode $N$-photon state $\mid \eta \rangle$ is defined
by the single tunable parameter $\eta$, which is given by $N$ times
the ratio of down-conversion pair amplitude $\gamma$ and squared
coherent amplitude $\alpha$.

We can now transform relation (\ref{eq:op}) into a non-linear
squeezing relation of the Schwinger parameters. The result
reads \cite{Ono07}
\begin{eqnarray}
\lefteqn{
\left(1+\left(\frac{1}{2}+\frac{\hat{J}_1}{N}\right) \eta \right)
\hat{J}_2 \mid \eta \rangle
=}
\nonumber \\ &&
i \left(1-\left(\frac{1}{2}+\frac{\hat{J}_1}{N}\right) \eta \right)
\hat{J}_3 \mid \eta \rangle.
\label{eq:nls}
\end{eqnarray}
It is possible to derive the approximate features of the quantum
state $\mid \eta \rangle$ by appropriate linearizations of this squeezing relation.

\section{Morning states: from linear squeezing to path entanglement}
\label{sec:am}

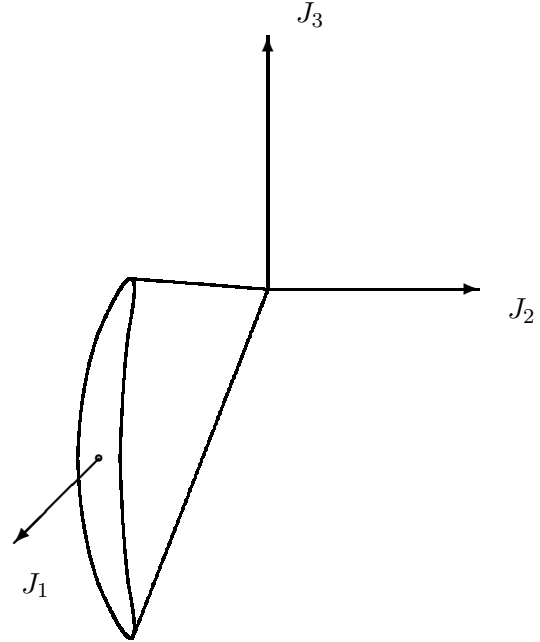
\begin{figure}[ht]
\setlength{\unitlength}{0.8pt}
\begin{picture}(300,320)


\thicklines

\bezier{100}(60,90)(61,125)(70,150)
\bezier{100}(70,150)(81,175)(85,175)
\bezier{100}(85,175)(89,175)(84,150)
\bezier{100}(84,150)(81,125)(80,90)
\bezier{100}(80,90)(81,55)(84,30)
\bezier{100}(84,30)(89,5)(85,5)
\bezier{100}(85,5)(81,5)(70,30)
\bezier{100}(70,30)(61,55)(60,90)

\bezier{100}(85,175)(124,172)(150,170)
\bezier{200}(86,5)(124,103)(150,170)

\put(150,170){\vector(0,1){120}}
\put(150,290){\makebox(40,20){$J_3$}}
\put(150,170){\vector(1,0){100}}
\put(250,150){\makebox(40,20){$J_2$}}
\put(70,90){\circle{3}}
\put(70,90){\vector(-1,-1){40}}
\put(20,20){\makebox(40,20){$J_1$}}

\end{picture}
\caption{\label{dawn}
Dawn state ($\eta<1$). The addition of a small amount of
down-converted photon pairs causes phase squeezing.
}
\end{figure}

For $\eta<1$, most of the photons originate from the laser light
input, so $\langle \hat{J}_1 \rangle \approx N/2$. The linearized
squeezing relation then represents a gradual redistribution of
quantum noise from $\Delta J_2$ to $\Delta J_3$. Specifically,
\begin{equation}
(1+\eta) \hat{J}_2 \mid \eta \rangle \approx
i (1-\eta) \hat{J}_3 \mid \eta \rangle,
\end{equation}
so the ratio of uncertainties in $\hat{J}_2$ and $\hat{J}_3$
is given by a squeezing factor of
\begin{equation}
\exp(2 r)=
\frac{\Delta J_3}{\Delta J_2} \approx \frac{1+\eta}{1-\eta}.
\end{equation}
A representation of this squeezed state is shown in fig. \ref{dawn}.
Since the state is still centered around the `eastern' horizon
at ${\bf J}=(N/2;0;0)$, we might call this the dawn state of our
quantum interference scheme.

At $\eta=1$, there is a transition between squeezing and
a non-classical superposition of two separate regions on
the ${\bf J}$-sphere. As we discuss in detail in \cite{Ono07},
this transition results in a maximal phase squeezing of
about $\delta \phi^2 = 1/N^{3/2}$ near $\eta=1$, corresponding
to the geometric mean of standard quantum limit and Heisenberg limit.
For $\eta>1$, the state splits into a superposition of two
${\bf J}$-vectors with opposite values of $J_3$, corresponding to
opposite intensity distributions between the two paths in the
interferometer.
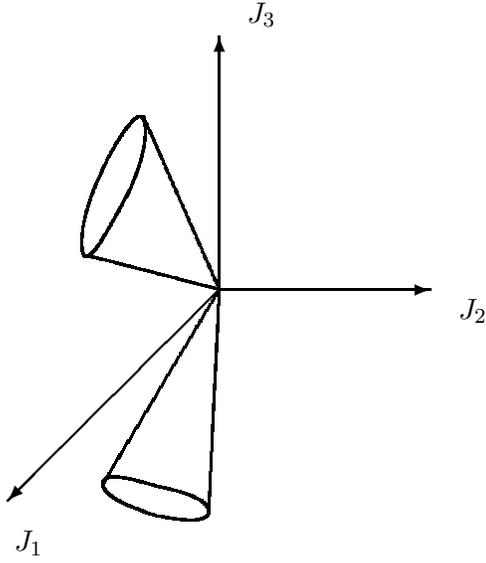
\begin{figure}[ht]
\setlength{\unitlength}{0.8pt}
\begin{picture}(300,320)
\thicklines

\bezier{100}(150,170)(120,120)(95,78)
\bezier{100}(150,170)(148,128)(145,65)

\bezier{100}(120,63)(145,58)(145,65)
\bezier{100}(145,65)(145,72)(120,77)
\bezier{100}(120,77)(95,85)(95,78)
\bezier{100}(95,78)(95,71)(120,63)

\bezier{100}(150,170)(132,211)(114,252)
\bezier{100}(150,170)(119,178)(88,186)

\bezier{100}(100,237)(115,262)(115,245)
\bezier{100}(115,245)(115,228)(100,203)
\bezier{100}(100,203)(85,175)(85,192)
\bezier{100}(85,192)(85,209)(100,237)


\put(150,170){\vector(0,1){120}}
\put(150,290){\makebox(40,20){$J_3$}}
\put(150,170){\vector(1,0){100}}
\put(250,150){\makebox(40,20){$J_2$}}
\put(150,170){\vector(-1,-1){100}}
\put(40,40){\makebox(40,20){$J_1$}}

\end{picture}

\vspace{-1.2cm}

\caption{\label{morning}
Morning state ($1<\eta<2$). The state describes a quantum
superposition of $\hat{J}_2$-squeezed states with squared
squeezing factors greater than two.
}
\end{figure}
The approximate ${\bf J}$-vectors of this superposition can be
found by noting that the classical limit of eq.(\ref{eq:nls}),
where operators are replaced by real numbers, requires
that $J_2=0$ and that either $J_3$ itself or the factor before
$J_3$ must be zero. For $\eta>1$, the classical limit
permits solutions with $J_3 \neq 0$. From these solutions,
the approximate average of $\hat{J}_1$ can be derived as
\begin{equation}
\langle \hat{J}_1 \rangle \approx \frac{N}{2}
\left( \frac{2}{\eta}-1 \right).
\label{eq:J1}
\end{equation}
In the classical limit, the two ${\bf J}$-vectors that solve
eq.(\ref{eq:nls}) have opposite $J_3$ values of
$\pm N \sqrt{\eta-1}/\eta$. The amount of squeezing
can then be determined by linearizing the squeezing
relation (\ref{eq:nls}) around the corresponding points in the
$J_1$-$J_3$ plane. The approximate $\hat{J}_2$-squeezing
thus obtained is described by
\begin{equation}
\Delta J_2^2 \approx \frac{\eta-1}{\eta} \frac{N}{4},
\end{equation}
indicating that the $\hat{J}_2$-squeezing continuously drops
back towards the shot noise limit of $N/4$ as $\eta$ increases.
A representation of the superposition state obtained at
$1<\eta<2$, just after the transition from squeezing to
quantum superpositions, is shown in fig. \ref{morning}.
Since the components of the quantum superposition are still
in the `eastern' half of the ${\bf J}$-sphere, we might call
this the morning state of our quantum interference scheme.

\section{Noon states: maximal path entanglement}

A special point is reached at $\eta=2$, where the
average value of $\hat{J}_1$ is zero. At this operating point,
the $N$-photon state is given by a superposition of two
squeezed states centered around the poles of the
${\bf J}$-sphere at $J_3=\pm N/2$. The squeezed noise
distribution at high $N$ is
approximately given by $\delta J_2^2=N/8$ and $\delta J_1^2=N/2$.
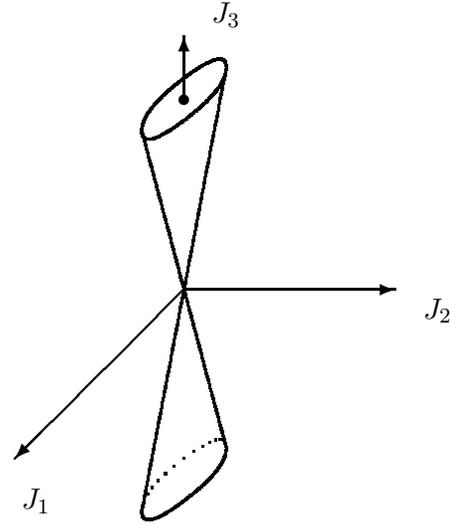
\begin{figure}[h]
\setlength{\unitlength}{0.8pt}


\begin{picture}(300,320)


\thicklines


\bezier{8}(150,90)(170,105)(170,95)
\bezier{100}(170,95)(170,85)(150,70)
\bezier{100}(150,70)(130,55)(130,65)
\bezier{8}(130,65)(130,75)(150,90)

\bezier{200}(170,95)(150,170)(130,245)
\bezier{200}(130,65)(150,170)(170,275)

\bezier{100}(150,270)(170,285)(170,275)
\bezier{100}(170,275)(170,265)(150,250)
\bezier{100}(150,250)(130,235)(130,245)
\bezier{100}(130,245)(130,255)(150,270)


\put(150,260){\vector(0,1){30}}
\put(150,260){\circle*{5}}
\put(150,290){\makebox(40,20){$J_3$}}
\put(150,170){\vector(1,0){100}}
\put(250,150){\makebox(40,20){$J_2$}}
\put(150,170){\vector(-1,-1){80}}
\put(60,60){\makebox(40,20){$J_1$}}

\end{picture}

\vspace{-1.2cm}

\caption{\label{noon}
Noon state ($\eta=2$). Since the quantum superposition between
states at $J_3=\pm N/2$ is squeezed by a squared factor of two,
the noon state fidelity at high $N$ is 94.3 \%.
}
\end{figure}
In the $\hat{J}_3$-basis describing the photon number distribution
between the paths inside the interferometer, this level of squeezing
requires only a small addition of the $\mid N-2,2 \rangle$ and the
$\mid 2,N-2 \rangle$ states to the ideal noon state,
\begin{eqnarray}
\mid \eta=2 \rangle &\approx& \left(\frac{2}{9}\right)^{1/4}
\Big(\mid N;0 \rangle + \frac{1}{3 \sqrt{2}} \mid N-2;2 \rangle
\nonumber \\ &&
+ \frac{1}{3 \sqrt{2}} \mid 2;N-2 \rangle +\mid 0;N \rangle \Big).
\end{eqnarray}
Thus the overlap between this approximate noon state and an ideal
noon state is given by
\begin{equation}
|\langle \mbox{NOON} \mid \eta=2 \rangle |^2
\approx \sqrt{\frac{8}{9}} \approx 0.943.
\end{equation}
At $\eta=2$, the interference between down-converted light and laser
light therefore results in a noon state with a fidelity of $94.3 \%$
in the limit of high photon number $N$ \cite{Hof07}.
Significantly, the generation of this state does not require any
post-selection conditions and results in a phase sensitivity that is
only slightly lower than the Heisenberg limit.
A representation of the approximate noon state state is shown in
fig. \ref{noon}. Since the upper component of the quantum
superposition has now reached the zenith, the noon state terminology
appropriately characterizes this state in the new context of our quantum interference scheme.

\section{Afternoon states: from cat-states to
kitten-states}

\begin{figure}[ht]
\setlength{\unitlength}{0.8pt}


\begin{picture}(320,320)


\thicklines

\bezier{100}(150,170)(161,196)(172,222)
\bezier{100}(150,170)(176,181)(202,192)

\bezier{100}(190,230)(210,230)(210,210)
\bezier{100}(210,210)(210,190)(190,190)
\bezier{100}(190,190)(170,190)(170,210)
\bezier{100}(170,210)(170,230)(190,230)

\bezier{100}(150,170)(155,135)(160,100)
\bezier{100}(150,170)(171,144)(192,118)

\bezier{100}(180,120)(200,120)(200,100)
\bezier{100}(200,100)(200,80)(180,80)
\bezier{100}(180,80)(160,80)(160,100)
\bezier{100}(160,100)(160,120)(180,120)


\put(150,170){\vector(0,1){120}}
\put(150,290){\makebox(40,20){$J_3$}}
\put(150,170){\vector(-1,0){100}}
\put(10,150){\makebox(40,20){$J_2$}}
\put(150,170){\vector(-1,1){60}}
\put(50,230){\makebox(40,20){$J_1$}}

\end{picture}

\vspace{-1.2cm}

\caption{\label{afternoon}
Afternoon state ($\eta>2$). $\langle \hat{J}_1 \rangle$
is negative and $\Delta J_2^2 \approx N/4$ corresponds to
shot noise.
}
\end{figure}
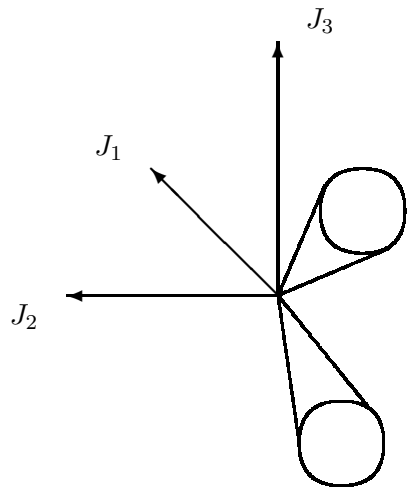

As $\eta$ increases beyond two, the squeezing effects quickly become
negligible and the quantum states can be described directly by a
superposition of two classically coherent states, where all of the
photons are in either one or the other of two non-orthogonal optical
modes. The ${\bf J}$-vectors of these two modes are approximately
given by the results derived in section \ref{sec:am}. In particular,
the average value of $\hat{J}_1$ is given by eq.(\ref{eq:J1}), and
the values of $\hat{J}_3$ are given by $J_3=\pm N \sqrt{\eta-1}/\eta$.
A representation of a state with $\eta>2$ is shown in
fig. \ref{afternoon}.
As $\eta$ increases, the two
${\bf J}$-vectors describing the non-classical superposition
approach each other, drawing closer to the `western' end of the
${\bf J}$-sphere. It therefore seems appropriate to refer to these
states as the afternoon states of our quantum interference scheme.

For sufficiently high $\eta$, the two ${\bf J}$-vectors will again
merge into a single state centered around $J_1=-N/2$. As the two
states begin to overlap, the quantum superposition has effects
on the uncertainty distributions that are very similar to
those described for the superpositions of coherent states in
the context of continuous variables \cite{Buz92}. In particular,
the states obtained for even $N$ are positive superpositions and
therefore exhibit $J_2$-squeezing, similar to the dawn states
of section \ref{sec:am}, while the states for odd $N$ are negative
superpositions and therefore exhibit a three fold increase in
$\Delta J_2^2$ as $\eta$ goes to infinity.
\begin{figure}[ht]
\setlength{\unitlength}{0.8pt}


\begin{picture}(320,320)


\thicklines

\bezier{100}(150,170)(168,175)(186,180)
\bezier{100}(150,170)(160,143)(170,116)

\bezier{100}(190,180)(210,180)(210,160)
\bezier{100}(210,160)(210,150)(200,143)
\bezier{100}(180,143)(170,150)(170,160)
\bezier{100}(170,160)(170,180)(190,180)

\bezier{100}(200,143)(195,140)(200,137)
\bezier{100}(180,143)(185,140)(180,137)

\bezier{100}(200,137)(210,130)(210,120)
\bezier{100}(210,120)(210,100)(190,100)
\bezier{100}(190,100)(170,100)(170,120)
\bezier{100}(170,120)(170,130)(180,137)


\put(150,170){\vector(0,1){120}}
\put(150,290){\makebox(40,20){$J_3$}}
\put(150,170){\vector(-1,0){100}}
\put(10,150){\makebox(40,20){$J_2$}}
\put(150,170){\vector(-1,1){60}}
\put(50,230){\makebox(40,20){$J_1$}}

\end{picture}

\vspace{-2.0cm}

\caption{\label{evening}
Evening state ($\eta>4 N$). The superposition merges at
$J_1=-N/2$, producing interference effects depending on
whether the total photon number $N$ is odd or even.
}

\end{figure}
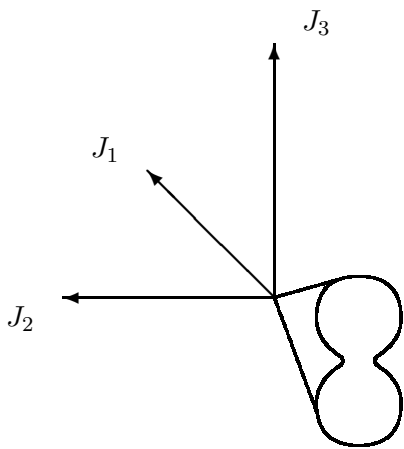
These phenomena become observable as soon as the $J_3$ values
of $\pm N \sqrt{\eta-1}/\eta$ become smaller than the quantum
fluctuations of $\sqrt{N}/2$. For high $N$, this happens around
$\eta = 4 N$. Fig. \ref{evening} illustrates the overlap of the
quantum state components at $\eta > 4 N$.
Since these states are close to the `western' horizon, we might
refer to them as the evening states of our quantum interference
scheme.

\section{Conclusions}
The interference between down-converted light and coherent laser
light provides a tunable source of non-classical $N$-photon states
that has the potential of greatly expanding the range of experimental
possibilities in quantum optics. As our analysis shows, the states
generated by this method cover both squeezing effects and extremely
non-classical superpositions, including a $94\%$ fidelity
approximation to the ideal noon state at $\eta=2$. When visualized
on the ${\bf J}$-sphere, the noon state appears as a quantum
superposition of a state at the $J_3=+N/2$ zenith and its mirror
image at $J_3=-N/2$. It is therefore tempting to identify the
terminology with the position of the ${\bf J}$-vector describing
the upper branch of the quantum superposition, leading to the
identification of morning, afternoon, and evening states.
Specifically, the linearly squeezed states described in
\cite{Ono07} can then be identified as morning states, while a
different kind of squeezing observed at the opposite (`western')
end of the ${\bf J}$-sphere can be identified as evening states.
It is thus possible to connect the noon state terminology with
an intuitive image of the quantum statistics associated with the
respective non-classical states.

\section*{Acknowledgements}
This work was partially supported by the CREST program of the
Japan Science and Technology Agency, JST, and the Grant-in-Aid
program of the Japanese Society for the Promotion of Science,
JSPS.


\begin{thebibliography}{xyz00}

\bibitem{Gio04}
A review of this problem is given in V. Giovannetti, S. Lloyd, and
L. Maccone, Science {\bf 306}, 1330 (2004).

\bibitem{Bow04}
D. Bowmeester, Nature (London) {\bf 429}, 139 (2004).


\bibitem{noonprop}
P. Kok, H. Lee, and J. P. Dowling,
Phys. Rev. A {\bf 65}, 052104 (2002);
J. Fiurasek, Phys. Rev. A {\bf 65}, 053818 (2002);
G.J. Pryde and A.G. White, Phys. Rev. A {\bf 68}, 052315 (2003);
H.F. Hofmann, Phys. Rev. A {\bf 70}, 023812 (2004).

\bibitem{noonexp}
P. Walther, J.-W. Pan, M. Aspelmeyer, R. Ursin, S. Gasparoni,
and A. Zeilinger, Nature (London) {\bf 429}, 158 (2004);
M.W. Mitchell, J.S. Lundeen, and A.M. Steinberger,
Nature (London) {\bf 429}, 161 (2004).

\bibitem{Hof07}
H.F. Hofmann and T. Ono, Phys. Rev. A {\bf 76}, 031806(R)
(2007).

\bibitem{QI}
Y. J. Lu and Z. Y. Ou, Phys. Rev. Lett. {\bf 88}, 023601 (2001);
F. Shafiei, P. Srinivasan, and Z. Y. Ou,
Phys. Rev. A {\bf 70}, 043803 (2004);
B. Liu and Z. Y. Ou,
Phys. Rev. A {\bf 74}, 035802 (2006).

\bibitem{Ono07}
T. Ono and H.F. Hofmann, e-print, arXiv:0708.2809 (2007).

\bibitem{Buz92}
V. Buzek, A. Vidiella-Barranco, and P.L. Knight, Phys. Rev. A
{\bf 45}, 6570 (1992).


\end{thebibliography}
\end{document}